# Nuclear symmetry energy: from nuclear extremes to neutron-star matter

**Pawel Danielewicz**

National Superconducting Cyclotron Laboratory and Department of Physics and Astronomy, Michigan State University, East Lansing, MI 48824, USA

Email: danielewicz@nscl.msu.edu

**Abstract:** Extrapolations from nuclei to neutron stars hinge on the symmetry term in nuclear binding formulas. The term describes reduction in the binding associated with neutron-proton (np) imbalance. Regrettably, binding formulas in the literature commonly lack an intrinsic consistency with regard to the symmetry term. Our elementary considerations determine the universal macroscopic limit for the term and predict its weakening in light nuclei due to emergence of the nuclear asymmetry-skin. Experimental systematic of isobaric-analogue states allows for a determination of the volume and surface coefficients within the macroscopic limit of the symmetry term, disregarding any reminder of a binding formula. The results are next exploited to constrain the dependence of symmetry energy on nuclear density, essential for the neutron-star predictions.

Availability of nuclear exotic beams [1], with an unusual neutron-proton (np) content, has spurred interest in the symmetry term [2] of nuclear binding-energy formulas [3,4], accounting for a reduction in the nuclear binding due to np-imbalance. Understanding of that term is, naturally, essential for extrapolations from nuclei to neutron stars [5,6]. A binding-energy formula [7] relies on treating the nucleus as a macroscopic object, which underscores the general phenomenological nature of nuclear science. We indicate that the basic [7] and many advanced binding formulas [8,9] are intrinsically inconsistent regarding the symmetry term. Elementary considerations [10] determine the universal macroscopic limit [10,11] for the symmetry term and predict that term's weakening in light nuclei, due to the emergence of nuclear asymmetry skin. The skin represents a relative displacement of neutron and proton distributions and is normally described in involved formalisms [12-15]. We point out that the experimental skin systematic restricts [10] the relative magnitude of symmetry coefficients describing changes in the nuclear macroscopic volume <u>and</u> surface energies with changing np-imbalance. The absolute magnitude of the volume and surface coefficients of the symmetry term may be constrained, disregarding any remainder of the binding formula, upon extending the np-interchange symmetry in the symmetry term, to the isospin symmetry [16], a rotational symmetry in the np-space. The specific constraints follow from the systematic of isobaric-analogue states [17] that are a consequence of the isospin symmetry. The results for the coefficients constrain next the dependence of the symmetry energy on nuclear density, required for neutron-star predictions [5,18-21].

A binding formula expresses the energy $E$ of a nucleus in terms of nucleon numbers, $E=E(A,Z)$. Here, $A$ and $Z$ are the net nuclear nucleon and proton numbers, respectively, and the neutron number is $N=A$-$Z$. The basic [7], termed Bethe-Weizsäcker [3,4] (BW), formula represents the energy as a sum of five terms only:

$$E = -a_V A + a_S A^{2/3} + a_C Z(Z-1)/A^{1/3} + a_A (N-Z)^2/A + \delta. \qquad (1)$$

The first negative and dominant term in equation (1), with the coefficient $a_V \gg 16$ MeV, is called the volume term. It represents a contribution to the energy from nuclear interior, for nucleons under the sole influence of attractive nuclear interactions. Proportionality to $A$ for that term reveals the short-range nature of nuclear interactions. The next term, $a_S A^{2/3}$, with $a_S \approx 18$ MeV, is the surface term proportional to the nuclear surface area, since the nuclear radii scale according to $R \approx r_0 A^{1/3}$. The constant so-called radius parameter $r_0 \approx 1.14$ fm represents an approximately constant (normal) density in the nuclear interiors, of $\rho_0 = 3/(4\pi r_0^3) \approx 0.16$ fm$^{-3}$. As nucleons close to the surface are subject to less attraction than in the interior, the binding in equation (1) gets reduced in proportion to the surface area, as compared to a system with ignorable boundaries. It follows that an increase in the nuclear surface area $\Sigma$ would reduce binding at an energetic cost per unit area, or surface tension, of $\sigma = \partial E / \partial \Sigma = a_S / 4\pi r_0^2$. The third term in the BW formula (1) is the Coulomb term, essentially representing the electrostatic energy of protons spread out uniformly over a spherical nuclear volume, with the coefficient correspondingly given by $a_C \approx (3/5)e^2/4\pi\varepsilon_0 r_0 \approx 0.7$ MeV. The fourth term in the formula, $a_A (N-Z)^2/A$, with $a_A \approx 21$ MeV, is commonly called a symmetry term although asymmetry would be a more suitable adjective. That term accounts for the binding, under the sole influence of nuclear interactions, being stronger for more symmetric nuclei, with $N \approx Z$, than for more asymmetric nuclei, with different $N$ and $Z$. That term is, in particular, related to a stronger attraction between neutrons and protons than between like nucleons. The symmetry with respect to the np interchange in the (a)symmetry term reflects the charge symmetry [7] of nuclear interactions. The final term in the formula (1) is the pairing term, $\delta = \pm a_P A^{-1/2}, 0$, with $a_P \approx 11$ MeV, describing the odd-even effect in nuclei. Relative to its average behaviour, the binding is a bit stronger for nuclei with even $N$ and $Z$ and it is weaker for nuclei with odd $N$ and $Z$. Only the pairing term is microscopic in the basic formula (1), with the term's $A$-dependence, though, phenomenological.

The simple binding formula (1) is surprisingly successful in practice. In fitting the five coefficients to the recent data set on energies of over 3100 nuclei [22], the rms deviation of the formula turns out to be just 3.6 MeV, for nuclear energies spanning the range of over 2000 MeV. Formula modifications [9], either motivated microscopically or ad hoc, may further improve the accuracy. Binding formulas play a major role in nuclear physics, as in allowing for assessing nuclear stability and in illuminating nuclear interactions. The volume term in the formula, e.g., indicates that, under the influence of nuclear interactions alone, the energy in uniform np-symmetric nuclear matter minimizes at the energy per nucleon of $-a_V \approx -16$ MeV and the density $r_0$. Efforts to reproduce that minimum microscopically led to the conclusion that three-nucleon interactions operate in nuclei [23].

Microscopic investigations at different densities $r$ indicate that the quadratic dependence of nuclear energy on relative asymmetry, exhibited in equation (1), remains valid for a variety of interactions and a range of $r$, down to the limit of neutron matter [2], $N>>Z$. Much uncertainty, though, concerns the coefficient in front of that quadratic energy dependence on asymmetry and, especially, concerns the coefficient's changes with $r$. The uncertainty hampers predictions for neutron stars, such as of star structure from hydrostatic equilibrium [2,5,18,20]. This is because the nuclear pressure, stabilizing the star, is proportional to the derivative of energy with respect to $r$ and because the pressure component related to the energy of symmetric matter alone is low close to the minimum of energy at $r_0$. The BW fit appears to yield, at $r_0$, for the coefficient $S(r)$ in energy per nucleon, $S(r_0) = a_A \approx 21$ MeV and, for the full energy per nucleon in pure neutron matter, $-a_V + a_A \approx 5$ MeV. Though differing by the concentration factor, the complete symmetry term in the net energy and the coefficient $S$ are habitually both called symmetry energy.

Common microscopic inclusions in a binding formula beyond (1), improving the formula performance, are those of the shell effects and of the diffuseness and exchange corrections to the Coulomb term. However, we shall indicate that the basic BW formula (1) is incomplete and inconsistent [10] already at the macroscopic level. Bringing about macroscopic consistency affects the conclusions on symmetry energy. The essence of the inconsistency is in the fact that the same binding cannot be subtracted twice from the dominant volume term in the formula (1).

The symmetry term in the BW formula, changing as $A$ when $N$ and $Z$ are scaled by one factor, exhibits a volume character such as the formula's leading term. Thus, formula (1) states that the interior contribution to the binding decreases, when magnitude of asymmetry increases. With less binding to compensate for in developing a surface for an asymmetric nucleus, the surface tension should then drop as compared to a symmetric nucleus. When accounting for the asymmetry dependence, the macroscopic surface properties gets completely specified when the tension is expressed in terms of the asymmetry chemical potential $\boldsymbol{m}_A = \partial E / \partial (N - Z)$, $\boldsymbol{s} = \boldsymbol{s}(\boldsymbol{m}_A)$, where nucleon numbers are treated as continuous variables for the smooth part of nuclear energy. With the n-p interchange symmetry present when the Coulomb interactions are disregarded, the nuclear asymmetry $N$-$Z$ vanishes for a vanishing $\boldsymbol{m}_A$. In general, the inverse Legendre-transformation relation for asymmetry, from $\boldsymbol{m}_A$ to $N$–$Z$, is

$$N - Z = \frac{\partial \Phi}{\partial \boldsymbol{m}_A}, \tag{2}$$

where $\Phi = \boldsymbol{m}_A(N - Z) - E$. Following the discussion above, for small asymmetries the tension should behave as

$$\boldsymbol{s}(\boldsymbol{m}_A) = \boldsymbol{s}_0 - \nu \boldsymbol{m}_A^2, \tag{3}$$

where $\boldsymbol{s}_0 = a_S / 4 \pi r_0^2$ and $\nu$ is some positive constant. We next explore the consequences of equation (3).

When the surface tension depends on asymmetry, so does the surface energy. The latter dependence produces an apparent conceptual paradox [10] directly following from equation (2): the nuclear interior cannot contain the full np-imbalance $N$-$Z$ of a nucleus! This paradox is resolved when considering a rigorous separation of the macroscopic quantities into the volume and surface components, as e.g. following Gibbs [24]. The component separation depends on the assumed surface location and that location is naturally set, for nuclei, by demanding a vanishing nucleon surface number. In the binary system, though, the neutron and proton surfaces may be displaced from each other. With no net nucleon number, the surface can carry then a net np-imbalance, with the overall nuclear imbalance separating into the volume and surface components: $N$-$Z$=$N_V$-$Z_V$+$N_S$-$Z_S$, where $N_S$=-$Z_S$. For a larger radius for neutrons than for protons, $N_S$ is positive.

Following equations (2) and (3), the surface imbalance $N_S$-$Z_S$ emerges linear in the asymmetry potential $\boldsymbol{m}_A$ and the surface energy then quadratic in the imbalance:

$$E_S = a_S A^{2/3} + a_A^S (N_S - Z_S)^2 / A^{2/3},$$ (4)

where we have introduced the surface symmetry coefficient $a_A^S$=$1/(16\pi r_0^2 \nu)$. A similar reasoning leads to the nuclear volume energy quadratic in the volume imbalance:

$$E_V = -a_V A + a_A^V (N_V - Z_V)^2 / A,$$ (5)

where the volume symmetry coefficient $a_A^V$ is now generally different from $a_A$ in the BW formula (1). The Coulomb interactions remain, for the moment, ignored. In the nuclear ground state, the net macroscopic energy $E_V$+$E_S$ should be minimal under the condition of a fixed net imbalance $N$-$Z$. Quadratic in the imbalance, the two energies, $E_V$ and $E_S$, are analogous to the electrostatic energies of charged capacitors, proportional to the capacitor charges squared divided by the capacitances. For the minimal energy of the surface and volume capacitor combination, the net imbalance partitions itself in proportion to the capacitances,

$$\frac{N_S - Z_S}{N_V - Z_V} = \frac{A^{2/3} / a_A^S}{A / a_A^V} = A^{-1/3} \frac{a_A^V}{a_A^S}.$$ (6)

The minimal net energy further emerges in terms of the net imbalance squared divided by the net capacitance,

$$E_V + E_S = -a_V A + a_S A^{2/3} + \frac{(N - Z)^2}{A / a_A^V + A^{2/3} / a_A^S}.$$ (7)

On adding the Coulomb and pairing contributions, we now get the modified binding formula:

$$E = -a_V A + a_S A^{2/3} + a_C Z(Z-1)/A^{1/3} + a_A(A)(N\text{-}Z)^2/\text{A} + \delta,$$ (8)

where

$$a_A(A) = \frac{a_A^V}{1 + A^{-1/3} a_A^V / a_A^S}.$$ (9)

The $A$-dependent symmetry coefficient here weakens for low mass numbers $A$, as more of the np-imbalance gets pushed out from the nuclear interior to surface. Whether the $A$-dependent symmetry coefficient may be replaced in heavy nuclei by $a_A^V$ depends on the ratio $a_A^V/a_A^S$. Of the two symmetry coefficients in the formula, only $a_A^V$ contributes to the energy of normal neutron matter.

Different radii for neutron and proton distributions, characteristic for surface asymmetry excess, are detected in nuclei (cf. references in [10]), *albeit* with some difficulty. The excess is referred to as asymmetry skin. For determination of the difference of radii, the distribution of chargeless neutrons needs to be probed and usually combining results from different probes of a nucleus is required. In establishing an empirical systematic of the radii difference, or skin size, exotic beams have been employed [25]. For stable nuclei, parity violations in electron scattering had been proposed for discerning the neutron radii [26]. Theoretically, in microscopic calculations of the symmetry skins, numerical experimentation has been employed [12,13] to assess the relation of radii difference to the characteristics of bulk matter including $S(\boldsymbol{r})$.

The macroscopic equation (6) indicates that data on surface asymmetry can directly constrain the nuclear ratio $a_A^S/a_A^V$. Two issues, however, must be resolved before arriving at any constraints. One is that the data pertaining to the surface excess, referred to above, are expressed in terms of the difference of rms radii for neutron and proton density-distributions, $<r^2>_n^{1/2}$-$<r^2>_p^{1/2}$, rather than in terms of $N_S$-$Z_S$. The geometric conversion between the two quantities is, though, straightforward as long as the surface diffuseness, characterising a particle distribution, is similar for neutrons and protons. The second issue is that, for heavy nuclei, the Coulomb effects can compete with the symmetry-energy effects. Against the minimal symmetry energy, the Coulomb forces try, on one hand, to push the proton relative to neutron surface out. On the other hand, the Coulomb forces try to polarize the nuclear interior. By minimizing a sum of the three energies, $E_V+E_S+E_C$, with the interior contribution to $E_C$ and with the symmetry contribution to $E_V$ cast in integral forms, the Coulomb effects get easily accounted for and an analytic result for the difference of radii is obtained [10,27]:

$$\frac{\langle r^2 \rangle_n^{1/2} - \langle r^2 \rangle_p^{1/2}}{\langle r^2 \rangle^{1/2}} = \frac{A}{6N\left(1 + A^{1/3} a_A^S / a_A^V\right)} \left[ \frac{N-Z}{Z} - \frac{a_C A^{2/3}\left(10/3 + A^{1/3} a_A^S / a_A^V\right)}{28 a_A^V} \right].$$

(10)

In equation (10), $<r^2>^{1/2} \approx 0.77 r_0 A^{1/3}$ is the overall nuclear rms radius. On the rhs, the Coulomb correction is proportional to $a_C$. Otherwise, the leading rhs term in (10), with various factors, represents the surface imbalance $N_S$-$Z_S$ from equation (6) converted, following geometry, to the fractional difference of radii. Notably, the minimized net energy leading to (10), as quadratic in the asymmetry changes around the minimum, is significantly less affected [10] by the Coulomb-symmetry energy competition than is the surface imbalance.

To see if the disregard of diffusivity or of other microscopic effects might hurt the predictive power of our skin-formula (10), we confront in figure 1 our predictions with a comprehensive set of nonrelativistic and relativistic mean-field calculations by Typel and Brown [12]. The figure shows the correlations between the skin of $^{208}$Pb and the skins of $^{138}$Ba and $^{132}$Sn. The symbols in the figure represent the mean-field calculations and the lines represent our formula. The results in the figure suggest an accuracy of 0.01 fm for our formula in representing the microscopic theory (while the Pb rms radius is 5.50 fm!).

Experimental errors for skin sizes are large because of the difficulties in measuring the neutron radii. The parity violation experiment [26] aims e.g. at a 1% error in the neutron radius of $^{208}$Pb, which transcribes onto a representative error of 0.06 fm either in the neutron radius or skin size. Given large errors, conclusions on the ratio $a_A^V/a_A^S$ can be aided by fitting the skin-result (10) to a multitude of nuclear data on the skins. In figure 2, we show the results of such a fit, with roughly horizontal lines, as one and two standard-deviation limits on the $a_A^V/a_A^S$ ratio, for an assumed value of $a_A^V$. In the fit, the skin data [10] for the following isotopes have been included: $^{12}$C, $^{20-23,25-31}$Na, $^{40,42,43,44,48}$Ca, $^{46,48,50}$Ti, $^{58,64}$Ni, $^{90}$Zr, $^{116,124}$Sn and $^{206-208}$Pb. While the

experimental uncertainties dominate the deduced uncertainty in $a_A{}^V/a_A{}^S$, in fitting here and later we allow for theoretical uncertainties. We estimate the latter from the excess of residuals for our fit, over those expected from the declared experimental uncertainties alone, in a strategy analogous to that of reference [8]. High values of the $a_A{}^V/a_A{}^S$ ratio, favoured by the fit results and displayed in figure 2, clearly invalidate [10] the $a_A (A)$-expansion in $A^{-1/3}$, see equation (9), underlying many binding formulas in use [8,9].

Constraining the symmetry coefficients directly, by fitting the modified binding formula to the binding-energy data, turns out to be treacherous [10], as conclusions on details of the different terms in the formula are interrelated. Thus, at a fixed $A$, there is a dependence on the asymmetry in the Coulomb term. There are further arguments for adding one more, Wigner, term [8,9], proportional to $|N-Z|$, to the formula. Moreover, the average relative asymmetry, $|N-Z|/A$, changes for known nuclei as $A$ changes. As a consequence, in an energy fit, the conclusions on asymmetry-dependent and asymmetry-independent terms in a formula become interrelated too. Short of going after all formula details [9,14], ideal for an absolute coefficient determination would be a seemingly impossible study of the symmetry term in isolation from the formula remainder.

A study in isolation is actually enabled by extending the charge symmetry of nuclear interactions to the charge independence, the invariance symmetry under rotations in the np space [7]. In an analogy to the spin-1/2 states, the nucleons form an isospin $T=1/2$ doublet, with the $z$-isospin projections of $T_z =\pm 1/2$ representing the proton and neutron, respectively. The nucleonic isospins couple to a net isospin for a given nuclear state. Due to the charge independence, within the set of nuclei of one $A$, or isobars, states can be found that are analogues of each other, representing different $T_z =(Z-N)/2$ components of one $T$-multiplet. Within the excitation spectrum of a nucleus with low $|N-Z|$, in particular, isobaric analogue states (IAS) can be found [7,17] representing the ground states of neighbouring nuclei with higher $|N-Z|$. Isospin conservation rules allow for the IAS identification in nuclear processes. A binding formula naturally generalizes [16] to the states of lowest energy for a given $T$ in a nucleus.

Under charge independence, nuclear contributions to the energy are isospin scalars. The symmetry energy, in particular, must be proportional to the nuclear isospin squared, with the symmetry term then generalized to:

$$a_A(A) \ (N\text{-}Z)^2/A \rightarrow \ a_A(A) \ 4T(T\text{+}1)/A \ . \tag{11}$$

In the ground state, the isospin $T$ takes on the lowest possible value: $T=|T_z|=|N\text{-}Z|/2$. In this modification, most [29] of the Wigner term gets absorbed into the symmetry energy and it represents there the effects of ground-state isospin-fluctuations. In the binding-formula generalization [16] to different $T$, the pairing term vanishes for half-integer $T$ (odd $A$), and is positive and negative, respectively, for odd and even $T$. With the above generalization, in the macroscopic limit, the excitation energy of an IAS, representing the ground state of a neighbouring nucleus, is proportional to $a_A(A)$, provided that either $A$ is odd or that the IAS and the ground state of the current nucleus are characterized by $T$ of the same evenness.

Next, we determine $a_A(A)$ for individual $A$ from the maximal measured IAS excitation energies [17] that are not biased by a pairing contribution, using

$$a_A(A) = A\mathbf{D}E/4\mathbf{D}[T(T\text{+}1)] \ , \tag{12}$$

where $\mathbf{D}$ stands for the difference between the IAS and ground-state quantity (see the caption for Fig. 3 for more details). With the expected average linear dependence of the inverse coefficient on $A^{-1/3}$, $(a_A(A))^{-1} = (a_A^V)^{-1} + (a_A^S)^{-1} A^{-1/3}$, we plot $(a_A(A))^{-1}$ from equation (12) as a function of $A^{-1/3}$ in figure 3. Indeed, an average, approximately linear, decrease with a decrease in $A^{-1/3}$ is observed. Scatter around the average behaviour is attributable to microscopic effects. A weighted linear fit to the data, produces $(a_A^V)^{-1}$ as an intersect of the line with the vertical axis, and $(a_A^S)^{-1}$ as a slope. In figure 2, we indicate the 1σ and 2σ contour lines in the $a_A^V\text{-}\ a_A^V/a_A^S$ plane for the linear fit. Here, the coefficient uncertainties are dominated by the theory limitations seen in figure 3.

A couple of comments on the coefficient determination from IAS are in order. The first comment concerns the role of the isospin-symmetry breaking Coulomb-term of the binding formula, within the procedure. Thus, the square of the net nuclear isospin may be represented as

$$T(T+1) = T_\perp^2 + T_z^2,$$ (13)

where $T_\perp$ stands for the isospin perpendicular to the direction in isospin space along which n and p point. The ground state and the IAS, in the determination of $a_A(A)$, differ in the $T_\perp^2$ value but not in the $T_z$ value. The isospin difference in equation (12) amounts then to $D[T(T+1)] = D[T_\perp^2]$. The Coulomb term, however, depends on $T_z$ but not on $T_\perp^2$. Correspondingly, the Coulomb term and part of the symmetry term proportional to $T_z^2$, engaged in the Coulomb-symmetry energy competition, drop out from the macroscopic energy difference for $a_A(A)$. The second comment concerns testing a potential impact of the surface curvature on the symmetry coefficient and on conclusions, in the view of the strong impact of the surface on the symmetry coefficient evident in figure 3. To test the potential impact, we resort to the Thomas-Fermi (TF) theory [28,10], with a nonlocal term in the energy density proportional to the density gradient squared. By adjusting the strength of the nonlocal term, a realistic diffuseness of the nuclear surface may be achieved. In [10], the TF theory was used to test equation (10). We fit the data in figure 3, adjusting the magnitude and density dependence of $S(\boldsymbol{r})$, within a polynomial parameterization for $S$. The best-fit results are shown with filled squares in figure 3, together with the result for infinite matter. While the best-fit TF results do not exactly fall along the best-fit line in figure 3, they do not quite exhibit the naïvely expected curvature effects either, in terms of a weakly parabolic dependence on the abscissa. The symmetry volume and surface coefficients for the best-fit TF theory, indicated in figure 2, end up, in fact, rather close to those from the fit with equation (9). Under a closer examination, some wavering of the TF results along the best-fit line, appears associated with an $A$-dependence of the surface diffuseness for the simple TF theory, stronger than demonstrated by available electron scattering data. Notably, the TF rms deviation from data is significantly larger than the deviation for equation (9).

Correspondingly, in figure 2, we provide error limits, associated with data deviations, only for the coefficients from equation (9), and we end up using the TF results only for assessing the magnitude and direction of a possible systematic bias for the results from equation (9).

With the skin and IAS constraints and the TF result in figure 2, we can now assess $a_A^V$ to be within the region of 30.0< $a_A^V$ <32.5 MeV and the coefficient ratio within 2.6< $a_A^V/a_A^S$ <3.0. This is at the edge of the broadly identified region following the global ground-state energy fit [10]. Improvements going in accuracy down to fractions of MeV in the coefficients require comprehensive microscopic considerations [30].

Nuclear surface augments the nuclear asymmetry capacitance, because the symmetry energy per nucleon $S$ drops with density $r$ in the surface. The ratio $a_A^V/a_A^S$, characterizing the surface-to-volume capacitance ratio, generally constrains [10] the shape of the density dependence, $S(r)/a_A^V$. When considering a continuous distribution of asymmetry capacitors, within the local density approximation, the $a_A^V/a_A^S$ ratio emerges, in fact, as an integral across the nuclear surface involving the shape of the density dependence:

$$\frac{a_A^V}{a_A^S} = \frac{3}{r_0} \int dr \frac{r(r)}{r_0} \left[ \frac{a_A^V}{S(r(r))} - 1 \right].$$  (14)

Using equation (14) as a guidance, we plot in figure 4 a correlation between the coefficient ratio $a_A^V/a_A^S$, deduced from the skin of $^{208}$Pb with equation (10), and the drop of the symmetry energy at half the normal density, $S(r_0/2)/a_A^V$, for a variety of effective interactions employed in structure calculations [13]. Given our limits on $a_A^V/a_A^S$, we find that we can realistically limit the drop of $S$ with density to: $S(r_0/2)=(0.58-0.69)a_A^V$. Within the commonly employed power parameterization $S=a_A^V(r/r_0)^g$, this implies limits on the power of density dependence to 0.54<$g$<0.77. With the $a^V$ value, our results for $a_A^V$ imply now the energy of $-a^V+a_A^V$=(14-17) MeV in $r_0$-neutron matter, in place of the naïve 5 MeV. While advanced approaches generally agree that the energy in $r_0$-neutron matter must be higher than that from the naïve consideration [8,23], the specific result here is arrived in a uniquely straightforward fashion.

The dependence of $S$ on $r$ impacts the pressures reached within a neutron star, at different $r$, and the star structure. Calculations [18] have established correlations, of different tightness, between the size of a star for a given mass and pressure values at different densities. Given our limits on $S(r)$ around $r_0$ and the predominance of symmetry pressure there, we can put limits on pressure in the neutron star,

$$P(r_0) = r_0^2 \, \mathrm{d}\, S/\mathrm{d}r \approx g \, r_0 \, a_A^{V} = (2.7\text{-}3.9) \text{ MeV/fm}^3 \ . \tag{15}$$

Following the scaling $RP^{1/4} \approx$ const of reference [18], for a representative neutron star, 1.4 times more massive than the Sun, we can then predict the radius [31] range of $R = (11.5\text{-}13.5)$ km. The variation of $S$ with $r$ that we found appears too slow to enable a sufficient high-$r$ proton-concentration, inside a neutron star, needed for the direct Urca process cooling [5,19,20]. The significant extrapolation [5,19] to supranormal $r$ needs, though, testing in central nuclear reactions [2,6].

A fit to binding-energy data [22] with our binding formula, with the symmetry energy modified following charge independence, equations (8), (9) and (11), yields symmetry-parameter values close to the region from other constraints, see figure 2. This is partly coincidental, as results of binding-energy fits are fragile, sensitive, following previous discussions, to secondary details in the formula. Irrespectively of those details, though, the introduction of surface symmetry energy greatly improves the performance of a binding formula [10,11] for low-mass highly asymmetric nuclei, such as those studied at the exotic beam facilities [1]. Figure 5 illustrates changes in the binding-energy residuals for $|N\text{-}Z|/A > 0.2$ nuclei, when switching from the best-fit standard formula to our last formula with $a_A^{V}/a_A^{S} = 2.8$ enforced. In the latter case, the parameter values are $a_V \approx 15.4$ MeV, $a_S \approx 16.9$ MeV, $a_C \approx 0.69$ MeV, $a_P \approx 11.6$ MeV and $a_A^{V} \approx 32.6$ MeV.

To summarize, considerations of consistency for the macroscopic nuclear energy function imply the emergence of nuclear asymmetry skins and weakening of the symmetry-energy term in light nuclei. Data on nuclear skins constrain the ratio of coefficients $a_A^{V}/a_A^{S}$ within the symmetry term whose form uniquely follows from the macroscopic considerations. Absolute magnitude of the coefficients is constrained by the mass dependence of excitation energies to IAS representing ground-states in the same isobaric multiplet. Using currently available data, we narrow the coefficients, following the macroscopic approach only, to within the range $30.0 < a_A^V < 32.5$ MeV and $2.6 < a_A^V/a_A^S < 3.0$. Looking at the vertical scales in figure 1, it is apparent that the parity violation measurement of lead neutron radius is not likely to shed new light on the symmetry energy. From equations (6) and (10), it is apparent that, at a given relative asymmetry, the relative skin size is largest for light nuclei. Further improved skin measurements for light highly asymmetric exotic nuclei [25] should be, on the other hand, beneficial in narrowing the uncertainty in $a_A^V/a_A^S$. Otherwise, microscopic theoretical investigations of the relative energies of isospin multiplets can yield insights into the symmetry energy, with the effects of charge invariance and charge symmetry [32] being of particular interest. The coefficient ratio in $a_A^V/a_A^S$ is related to the shape of the density dependence of symmetry energy at subnormal densities. Within the common power parameterization of the density dependence, $S/a_A^V = (r/r_0)^g$, at moderately subnormal densities, $0.5r_0 < r < r_0$, we find $0.54 < g < 0.77$. This has a further a further bearing on the pressure in neutron matter, $P(r_0) = (2.7$-$3.9)$ MeV/fm$^3$ and on properties of neutron stars.

**Acknowlegements**

I thank Peter Möller for discussions and Wladek Swiatecki and Betty Tsang for suggestions regarding the manuscript. This work was supported by the National Science Foundation under Grant PHY-0245009.

**Figures**

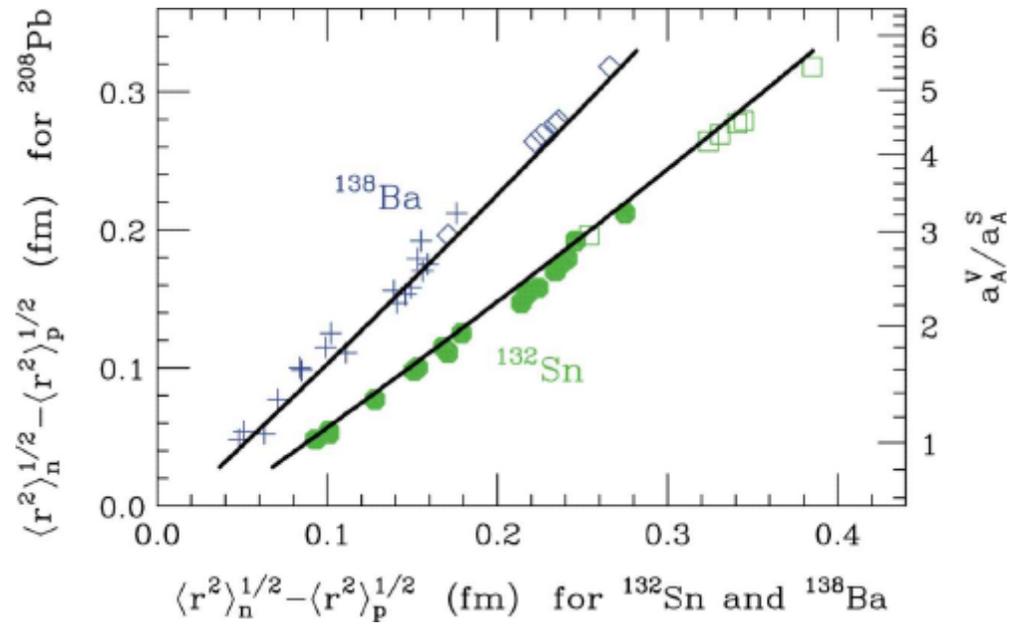

**Figure 1.** Asymmetry-skin correlation between the $^{138}$Ba, $^{132}$Sn and $^{208}$Pb nuclei. The skins are quantified with the difference of neutron and proton rms radii. Symbols represent the results [12] from nonrelativistic Hartree-Fock (crosses and filled circles) and relativistic Hartree (diamonds and squares) calculations, for a variety of effective Lagrangians and Hamiltonians. The lines represent our analytic formula (10), from minimizing the combination of macroscopic symmetry and Coulomb energies. The rhs scale of the figure shows the symmetry coefficient ratio $a_A^V/a_A^S$ for our formula. Given the weak sensitivity of skin size to $a_A^V$ in separation from $a_A^S$, solely through $a_C/a_A^V$ in the Coulomb term in (10), in obtaining the rhs scale we set $a_A^V$, for a given $a_A^V/a_A^S$, by insisting on the consistency with the BW coefficient, $a_A(200) \approx 21$ MeV.

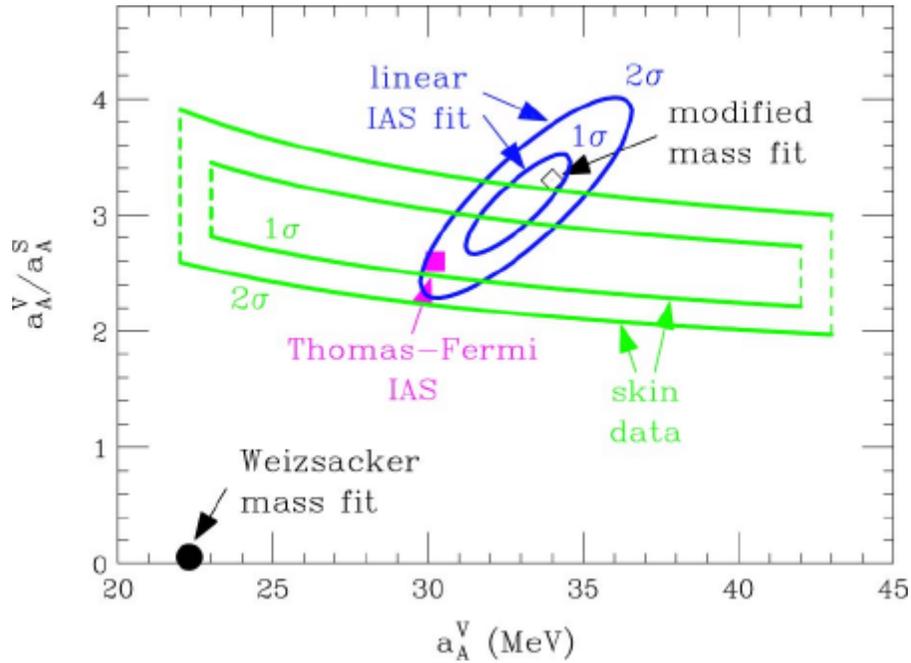

**Figure 2.** Results for the symmetry parameters in the plane of $a_A^V/a_A^S$ vs $a_A^V$. Slant lines represent the $1\sigma$ and $2\sigma$ constraints on $a_A^V/a_A^S$ at fixed $a_A^V$, from the fit of our skin formula to the measurements quoted in reference [10]. Oval contours represent the constraints on $a_A^V/a_A^S$ and $a_A^V$, from fitting the linear dependence on $A^{-1/3}$, $(a_A(A))^{-1} = (a_A^V)^{-1} + (a_A^S)^{-1} A^{-1/3}$, to the mass-dependent symmetry-coefficient values from the IAS excitation energies [17], shown in figure 3. The filled square represents the symmetry parameters obtained when fitting $a_A(A)$ from IAS within the simple Thomas-Fermi theory [10,28]. The filled circle and diamond represent, finally, the symmetry parameters from the best fit to the nuclear data [22], respectively, for the standard BW formula (1) and for our final binding formula (equations (8), (9) and (11)) including the effects of isospin fluctuations.

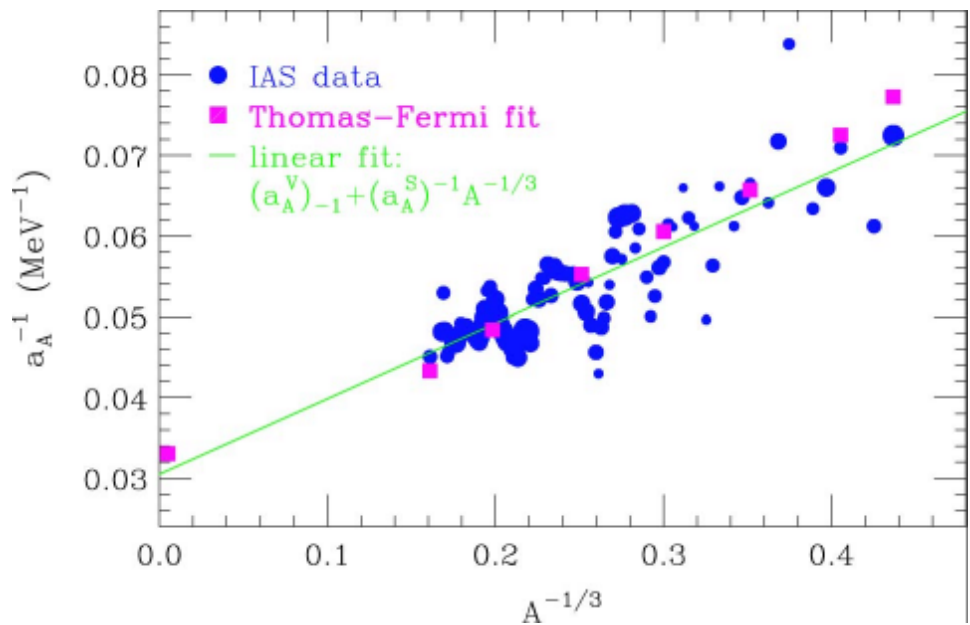

**Figure 3.** Inverse of the *A*-dependent symmetry-coefficient as a function of $A^{-1/3}$. Filled circles indicate inverse coefficient values obtained, using equation (12), from extremal measured excitation energies of the IAS [17] representing ground states of neighbouring nuclei. For an even *A*, the represented neighbouring nucleus was required to be of the same *Z*-evenness as the nucleus with the IAS. The evenness requirement was to preclude a pairing contribution to the energy difference interpreted, in the macroscopic limit, as associated with the symmetry energy alone. In some cases, an IAS representing the ground state of a neighbouring nucleus was not known, but an IAS of some low-lying excited state was. If the latter state's excitation energy did not exceed ~1 MeV, the energy of the known IAS was used and corrected for the excitation energy. Results from different nuclei of the same *A* were combined. Size of the symbols in the figure is proportional to the weight $4\boldsymbol{D}[T(T+1)]/A$ in the determination of $(a_A(A))^{-1}$, or to the weight combination. The larger the weight (generally of the order of 1) the more likely is the suppression of fluctuating microscopic contributions to the energy. Regarding those, the oscillating

pattern in $a_A^{-1}$ in the figure is characteristic for shell effects. The straight line in the figure represents the optimal weighted linear-fit to the data for $a_A^{-1}$. The filled squares represent the optimal weighted Thomas-Fermi fit to those data.

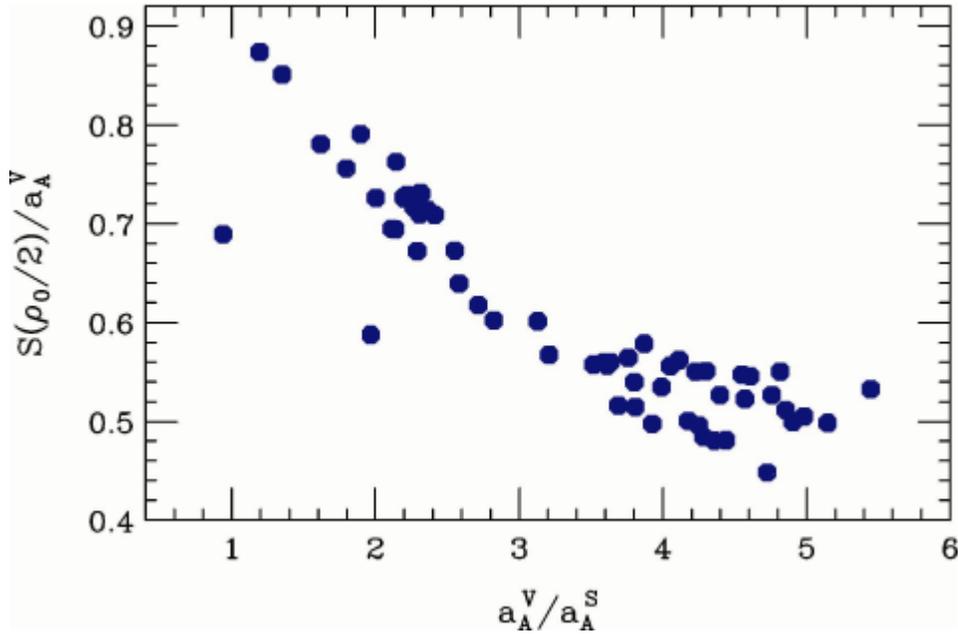

**Figure 4.** Correlation between the symmetry-coefficient ratio $a_A^V/a_A^S$ and between the reduction in symmetry energy at half of normal density, relative to the normal, $S(r_0/2)/a_A^V$, for a variety of mean-field models explored by Furnstahl [13]. The coefficient ratio for the bottom scale was obtained from the $^{208}$Pb skins for the models following equation (10), cf. figure 1. (The apparent small but systematic Coulomb discrepancy for the formula, reported for comparisons [10] with Thomas-Fermi calculations, turned out to be related to an insufficient accuracy of those calculations.) A smooth-curve fit to the results in the figure produces an uncertainty range of $0.58 < S(r_0/2)/a_A^V < 0.69$, for $2.6 < a_A^V/a_A^S < 3.0$.

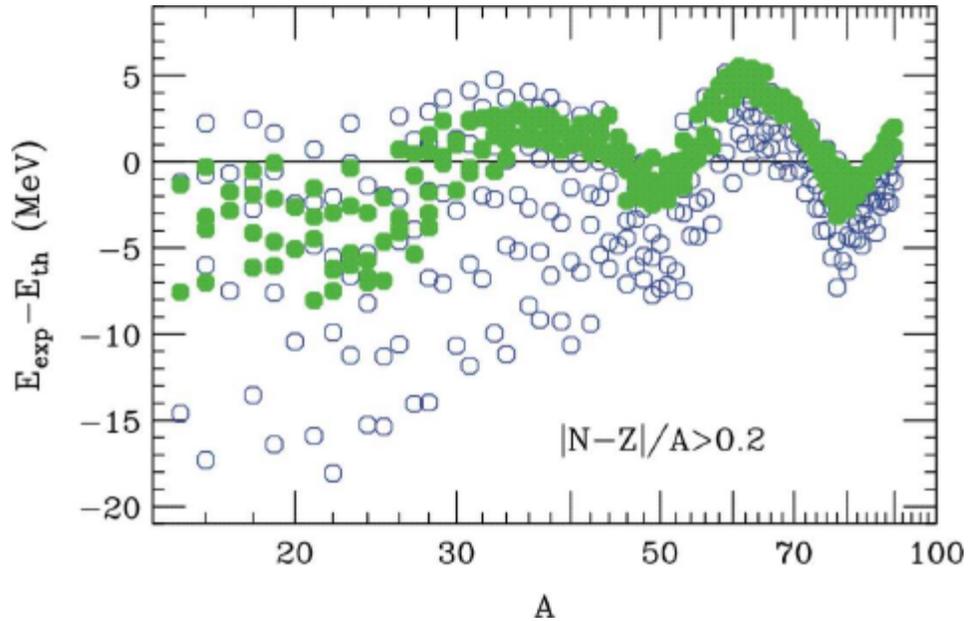

**Figure 5.** Residuals of binding-formula fits to known ground-state nuclear energies [22], shown for the low-mass ($A$<90) region of asymmetric nuclei, |$N$-$Z$|/$A$>0.2. Open and closed symbols represent, respectively, the residuals when following the standard formula and the formula with the symmetry term of the form $a_A^V$|$N$-$Z$| (|$N$-$Z$|+2)/($A$+$A^{2/3}$ $a_A^V/a_A^S$) with $a_A^V/a_A^S$ =2.8 enforced. (The number of fitted parameters is then the same for the two fits.) Observed oscillations in the residuals are generally characteristic for shell effects. Some scatter of the residuals for the modified formula, persisting at low mass numbers, is due to unaccounted competition between the symmetry and Coulomb energies [10]. When switching between the formulas, for the $A$<50 and |$N$-$Z$|/$A$>0.2 region, the rms deviation from data drops from 6.6 MeV down to 3.0 MeV, becoming close to the rms deviation for all nuclei of 2.7 MeV.